\documentclass[%
 reprint,
 amsmath,amssymb,
 aps,
 prl,
]{revtex4-2}

\usepackage{graphicx}
\usepackage{dcolumn}
\usepackage{bm}
\usepackage{xcolor}
\usepackage{comment}

\begin{document}

\preprint{APS/123-QED}

\title{Scaling of high-Rayleigh-number convection based on internal convective boundary layer}
\thanks{ctong@clemson.edu}%

\author{Chenning Tong}
 \affiliation{Department of mechanical Engineering, Clemson University, Clemson, SC 29631} 




\date{\today}

\begin{abstract}
  We propose a phenomenological model for thermal convection at  high Rayleigh numbers.
  It hypothesizes existence of a high-Reynolds-number turbulent boundary layer
  near each horizontal plate, which is shown to be convective.
  The convective logarithmic friction law of Tong and Ding (2020) is used to relate the large-scale
  velocity to the induced friction velocity. 
  The predicted scaling relations of the Nusselt ($Nu$)  and Reynolds numbers ($Re$)
  on the Rayleigh number ($Ra$)  do not have a power law form, each being a single function, suggesting a single flow regime with no transition, instead
  of multiple regimes.  However, the predicted $Nu$ and $Re$ scaling is close to $Ra^{1/3}$ and $Ra^{4/9}$
  for $Ra \sim 10^9$ to $10^{17}$, consistent with direct numerical simulation (DNS) results up to $Ra\sim 10^{15}$.
  For $Ra<10^9$, the model also correctly predicts the deviations observed in DNS from the above power law scaling.
  The model predicts deviations from $Ra^{1/3}$ beyond $Ra \sim 10^{17}$ with the local scaling exponent approaching $1/2$,
  which would likely require data at $Ra>10^{18}$ to verify.
\end{abstract}

\maketitle


\noindent \textit{Introduction } Rayleigh-B\'{e}nard convection has been a flow of great interest for decades
\cite{Malkus1954,Priestley1959,Kraichnan62,Howard1963,Toomre1977,Krishnamurti1981,Castaing1989,Shraiman1990,She1989,Wu1992,Siggia94,Ciliberto1996,
  Shen1996,Chavanne1996,Chavanne1997,Cioni1997,Constantin1999,Glazier1999,Ashkenazi1999,Grossmann2000,Niemela2000,Niemela2006,Iyer2020,Ahlers2022,Lindborg2023}.
The central issue in understanding its physics
is the scaling of the Nusselt number, $Nu=QD/(\alpha \Delta T)$, with the Rayleigh number, $Ra=g \Delta T D^3/(T \nu \alpha)$,
where $Q$, $T$, $\Delta T$, $D$, $g$,  $\alpha$, $\nu$ are the mean temperature flux, the mean temperature, the temperature difference between
the hot and cold plates,
the distance between the plates, the gravitational acceleration, the thermal diffusivity, and the viscosity, respectively.
The scaling is generally considered to be power law. Dimensional arguments and
marginal stability analysis results in $Ra^{1/3}$ \cite{Malkus1954,Priestley1959,Siggia94}. 
On the other hand, Kraichnan  \cite{Kraichnan62} predicted using mixing-length arguments that the $Ra^{1/3}$ scaling only valid for moderate $Re$
and predicted $Ra^{1/2}/(\ln Ra)^{3/2}$ for very large $Ra$, which is so called the ``ultimate'' regime.
While available data have not shown a transition to the ultimate regime \cite{Glazier1999,Niemela2000,Niemela2006,Iyer2020},
increasingly casting doubt on its existence,
there is insufficient evidence to rule out the regime.

In the present work, we propose a phenomenological model, focusing on predicting the $Nu$-$Ra$ and $Re$-$Ra$ scaling, and  only considering
the case of unity Prandtl number.
We hypothesize the existence of an internal turbulent boundary layer near each plate,
as done by Kraichnan  \cite{Kraichnan62}.
However, we show that the boundary layer is not a classical boundary layer, but a convective one.
We then use the model derive scaling relations.

\noindent \textit{The turbulent boundary layer} Our justification for hypothesizing the existence of such a boundary layer, which, to our best knowledge,
has not been put forward previously, is as follows.
At a distance $z$ from the surface, the vertical velocities of large-scale eddies 
of scales $1/k \gg z $ are blocked by the wall, resulting in quasi-two-dimensional motions, where $k$
is the wavenumber of the eddies. Therefore, the nonlinear interactions among such eddies are diminished.
Only eddies of scales $1/k\sim z$, which have three-dimensional motions at $z$, can have strong interactions with
the large eddies and can produce turbulent shear stress to damp their motions.
As a result, the horizontal velocities of the large eddies will extend toward the surface until they are damped at a height $\delta_o$ by 
 the eddies of scales $1/k\sim \delta_o$. The shear stress can be produced by the vertical velocity fluctuations
of eddies generated by the large-scale shear motions or by other eddies that are independent of the large-scale eddies (e.g.,
generated by buoyancy). In both cases, a
flow structure of thickness $\delta_o$ that resembles a turbulent boundary layer exists between the large-scale horizontal velocity and the surface.
This boundary layer has a horizontal length scale of order $D$ and a time scale of $D/w_*$, where $w_*=((g/T)QD)^{1/3}$ is large-scale convection velocity.

 To understand the structure of the boundary layer, we obtain the leading-order momentum balance in the outer layer as between the advection
 and the shear stress gradient
\begin{equation} \label{momentum}
\frac{w_*^2}{D} \sim \frac{v_*^2}{\delta_o},
\end{equation}
where $\delta_o$ and $v_*$ are the thickness of the boundary layer and the induced friction velocity respectively.
Note that while all scaling relations are based on fluctuating variables averaged over a horizontal scale of $D$, which are still random variables,
rather than ensemble averages, we expect the relations  to be valid in an average sense. 
The convection velocity in the outer layer  of the boundary layer scales as  $u_f \sim ((g/T)Q\delta_o)^{1/3}$. The ratio
\begin{equation}
\frac{u_f}{w_*} \sim (\frac{\delta_o}{D})^{1/3} \gg \frac{v_*}{w_*} \sim (\frac{\delta_o}{D})^{1/2},
\end{equation}
indicating that the vertical fluctuating velocity in the outer layer is dominated by convective eddies of scale $\delta_o$.
Therefore, the internal boundary layer is a convective boundary layer, rather a shear dominated one.
This is a key difference between the present work and that of Kraichnan \cite{Kraichnan62}.

Sufficiently close to the surface, the velocity fluctuations are dominated by eddies generated by the large-scale shear.
Therefore, the boundary layer structure is similar to that of the convective atmospheric boundary layer, and has a three-layer structure.
As noted above, the outer layer is buoyancy dominated. There are two inner layers. The inner-outer layer
scales with the friction velocity and the Obukhov length $L=-v_*^3/(\kappa gQ/T)$,
above which buoyancy production dominates (\cite{Obukhov49}), whereas below which shear production dominates, where $\kappa$ is the von K\'{a}m\'{a}n constant.
The inner-inner layer scales with the friction velocity and the viscous length scale $\delta_\nu=\nu/v_*$. Matching between the inner-outer and inner-inner
layer results in the log law.

For the convective boundary layer,  the convection logarithmic friction law derived using matched asymptotic expansions
by Tong and Ding \cite{TD19b}, when extended to a smooth surface, is
\begin{equation}
\frac{w_*}{v_*}=\frac{1}{\kappa}\ln \frac{-L}{\delta_\nu}+C,
\end{equation}
where $-L$ appears as the outer length instead of the boundary layer thickness in the logarithmic friction law for the zero-pressure-gradient boundary layer.
An intuitive way to understand the appearance of $-L$ is that
above the height $-L$ convection enhances the exchange of horizontal momentum, thus transporting fluid with velocities of order $w_*$ down to $z\sim -L$.
Note that the velocity scale for the viscous sublayer
 due to the local convective eddies (without considering the shear induced by the large convective eddies)
  would be  $v_\nu=((g/T)Q\delta_\nu)^{1/3}$. It then follows that $v_*/v_\nu=(v_*/w_*)^{4/3}(w_*D/\nu) \sim  Re/ (\ln Re)^{4/3} \gg 1$.
At sufficiently high $Re$ (and therefore $Ra$), $v_* \gg v_\nu$, 
indicating that near the surface the shear generated turbulence indeed dominates, and that the viscous length is $\delta_\nu$.  The convective logarithmic
friction law leads to different scaling than that of Kraichnan \cite{Kraichnan62}, which used
the friction law for neutral boundary layers.
The analogous logarithmic  heat transfer law is
\begin{equation}
\frac{\Delta T}{2T_*}=\frac{1}{\kappa}\ln \frac{-L}{\delta_\nu}+C,
\end{equation}
where $T_*=Q/v_*$ is the temperature scale near the surface.


\noindent \textit{Scaling relations}
The friction law, the heat transfer law, and the expression for $L$ determine the scaling of the convective boundary layer. The scaling of the Nusselt number
 and Reynolds number with
the Rayleigh number  can be obtained from these relations. 
To this end, we use their symmetry properties, specifically their Lie dilation group to obtain the scaling.
The group transformation can be written as
\begin{align} \label{eq_dilation1}
\begin{split}
 & \tilde{\nu}=e^{da}\nu,\ \tilde{v_*}=e^{db}v_*, \ \tilde{L}=e^{dc}L,  \ \tilde{Q}=Q,\  \tilde{D}=D, \\ 
  &   \tilde{w_*}=w_*, \ \widetilde{{\Delta T}}=e^{df}{\Delta T}, \tilde{T_*}=e^{dg} \ \widetilde{{Ra}}=e^{dh}{Ra}, \\
  & \widetilde{{Nu}}=e^{di}{Nu}, \ \widetilde{{Re}}=e^{dj}{Re}.
\end{split}
\end{align}
where  $da$, $db$ etc., are the differential group parameters. For the equations to be invariant the exponents must satisfy  $dc=3db$, $dg=-db$, $df=-2db$,
$dh=-2db-2da$, $di=2db-da$, and $dj=-da$.
Furthermore, writing the friction law (3) in terms of the tilde variables, the transformed equation is
\begin{equation}
\frac{w_*}{v_*}e^{-db}=\frac{1}{\kappa}\ln \{\frac{-Lv_*}{\nu}e^{4db-da}\}+C,
\end{equation}
which can be written as
\begin{equation}
\frac{w_*}{v_*}(1-db)=\frac{1}{\kappa}\{\ln \{\frac{-Lv_*}{\nu}+4db-da\}+C.
\end{equation}
Thus
\begin{equation}
-\kappa \frac{w_*}{v_*}db=4db-da,
\end{equation}
resulting in
\begin{equation}
da=(\kappa \frac{w_*}{v_*}+4)db.
\end{equation}

  From the group parameters we 
obtain the characteristic equations of the group
\begin{align} 
\begin{split}
 & \frac{dv_*}{v_*}=\frac{-dRa}{2Ra(\kappa\frac{w_*}{v_*}+5)}=\frac{-dNu}{Nu(\kappa\frac{w_*}{v_*}+2)}=\frac{-dRe}{Re(\kappa\frac{w_*}{v_*}+4)} \\
\label{eq_character_outer}
\end{split}
\end{align}
From the first and second terms we obtain (after non-dimensionalizing $v_*$ by $w_*$)
\begin{equation} \label{eq_Ra}
Ra \sim ({w_*}/{v_*})^{10}e^{2\kappa w_*/v_*}.
\end{equation}
Similarly we obtain
\begin{equation} \label{eq_Nu}
Nu \sim ({w_*}/{v_*})^{2}e^{\kappa w_*/v_*}, \ \ Re \sim ({w_*}/{v_*})^{4}e^{\kappa w_*/v_*}.
\end{equation}
Equations (\ref{eq_Ra}) and (\ref{eq_Nu}) are functions of $w_*/v_*$ which can be used as a parameter to obtain the scaling of $Nu$ and $Re$ on $Ra$.
From these equations we can also obtain a simple relation
\begin{equation}
Re^3 \sim RaNu.
\end{equation}

The non-dimensional coefficients in (\ref{eq_Ra}) and (\ref{eq_Nu}) can be obtained from the convective friction law. We can write
\begin{equation}
\frac{w_*}{v_*}=\frac{1}{\kappa}\ln \{\frac{-L}{D} \frac{v_*D}{\nu}\}+C=\frac{1}{\kappa}\{\ln (\frac{v_*}{w_*})^4 + \ln\frac{Re}{\kappa}\}+C.
\end{equation}
Thus
\begin{equation} \label{}
 Re =\kappa e^{-C} ({w_*}/{v_*})^{4}e^{\kappa w_*/v_*},
\end{equation}
\begin{equation} \label{eq_rac}
Ra= 2\kappa^2 e^{-2C} ({w_*}/{v_*})^{10}e^{2\kappa w_*/v_*}, 
\end{equation}
\begin{equation} \label{eq}
 \ Nu =\frac{1}{2}\kappa e^{-C} ({w_*}/{v_*})^{2}e^{\kappa w_*/v_*}.
\end{equation}
Equations (\ref{momentum}) and (\ref{eq_rac})  also provides an implicit relationship on the dependence of $\delta_o/D$ on $Ra$.

\vspace{6pt}

\begin{figure}
\centering
\includegraphics[width=3in,height=2.0in]{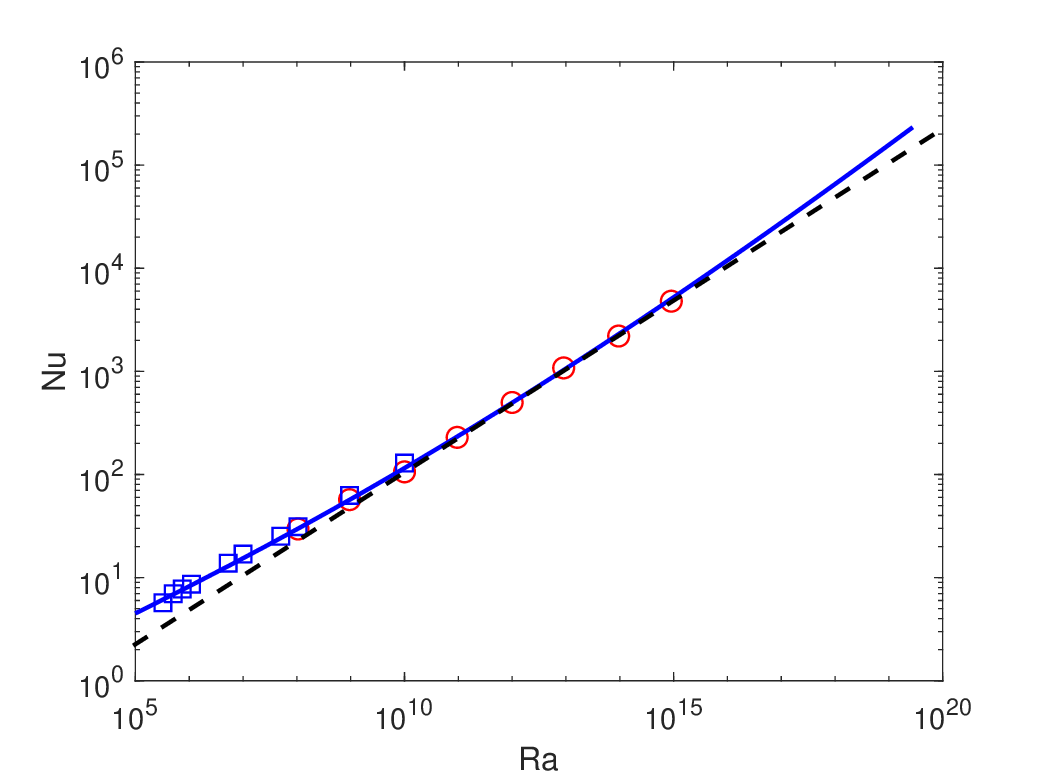}
\caption{$Nu$ vs.~$Ra$. DNS data from Iyer et al.~(2020). Circles: Pr=1.0; Squares: Pr=0.7); Solid line: Theoretical prediction. The dashed line has
  a slope of $1/3$.}
\vspace{.0in}
\label{fig1}
\end{figure}
\begin{figure}
\centering
\includegraphics[width=3.0in,height=2.0in]{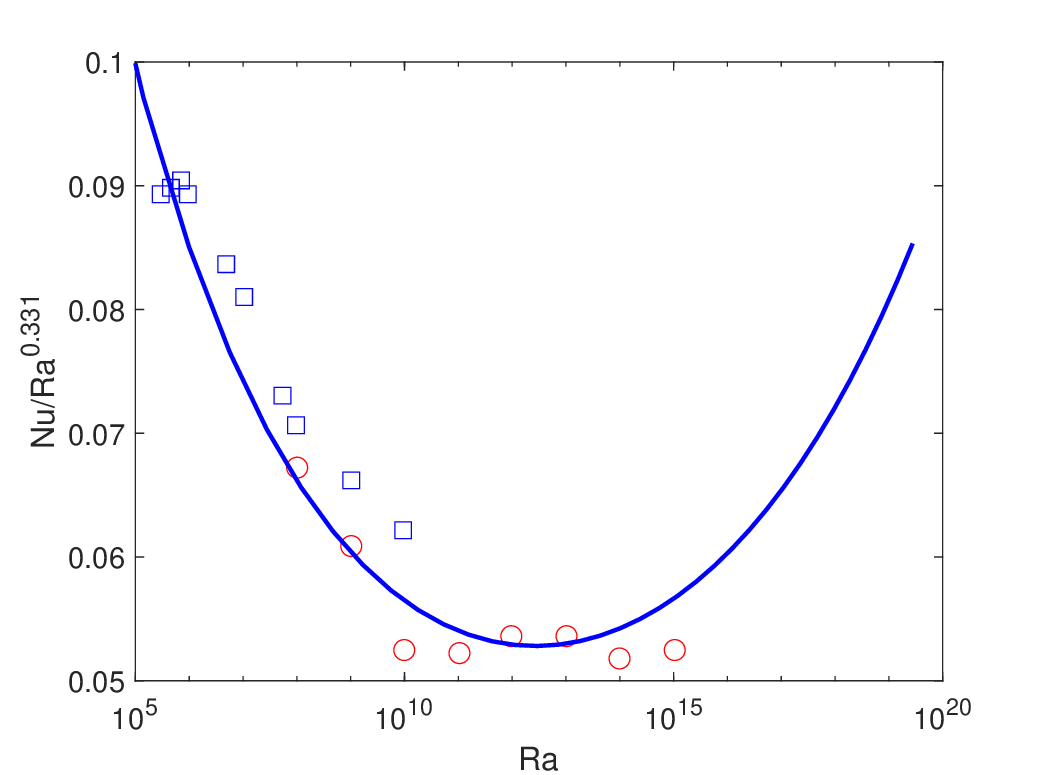}
\caption{$Nu/Ra^{0.331}$ vs.~$Ra$. Legends same as in Fig.~1.}
\label{fig2}
\end{figure}

The predicted $Nu$-$Ra$ scaling is shown in Fig.~1. Contrary to the previous predictions \cite{Malkus1954,Priestley1959,Kraichnan62,Siggia94},
the scaling is not in the form of power law. The
scaling relation has a single functional form, indicating that there is only one flow regime at high $Ra$, rather than two regimes as predicted
by Kraichnan \cite{Kraichnan62}.
From Eqs.~(\ref{momentum}) and (\ref{eq_rac}) we obtain $Nu\sim (w_*/v_*)^{-3}Ra^{1/2}$, which has some resemblance to the prediction
of Kraichnan \cite{Kraichnan62}, but the correction is not a simple logarithmic one. 
Figure 1 shows that our prediction is close to $Nu \sim Ra^{1/3}$ for $Ra \sim 10^9$ to $10^{17}$. For much higher $Ra$, due to the slow variations
of $w_*/v_*$, the scaling exponent will approach $1/2$. 
However, our model predicts no change in the leading-order physics as $Ra \rightarrow \infty$, 
indicating that the approximate $Ra^{1/3}$ scaling and the $Ra^{1/2}$ scaling, which if actually exists, belong to the same flow regime.
Therefore,  our model predicts  no further transition at higher $Ra$, contrary to the prediction of Kraichnan \cite{Kraichnan62},
although the local scaling exponent continuously increases toward $1/2$.

Figure 1 also compares the predicted $Nu$-$Ra$ scaling is also compared with DNS data of Iyer et al.~\cite{Iyer2020}. 
The von K\'{a}rm\'{a}n constant is chosen to be $\kappa=0.41$ (We also tried $\kappa=0.39$ and found that the results are not sensitive to small changes
in the value of $\kappa$). The additive constant in the
friction law $C=0.25$ is determined by fitting the theoretical prediction to the data. Note that $C$ only contributes to the constants of proportionality
and does not change the scaling. The prediction compares very well with the data.
The approximate $Ra^{1/3}$ scaling for $Ra\sim 10^9$ to $10^{15}$ as well as the deviation from the power law scaling for $Ra<10^9$,
the latter having not been previously explained, are both well predicted.
 Figure 2 further compares $Nu/Ra^{0.331}$ vs.~$Ra$. The rise for $Ra<10^{10}$ is also quite well predicted, suggesting that these
 $Ra$ values can also be considered as in the high $Ra$ range. Given the agreement between our prediction and the data, and the fact that the
 seemingly different scaling behaviors above and below $Ra \sim 10^9$ are predicted by our model to belong to a single high $Ra$ regime
 with the same leading-order physics,
 We might have good reason to expect the predicted $Nu$ scaling to hold at higher $Ra$ ($>10^{15}$).
 However, to distinguish the prediction from the 
 $Ra^{1/3}$ scaling beyond $Ra=10^{15}$ would likely require data at $Ra>10^{18}$.

\vspace{6pt}

\noindent \textit{Conclusions}
We proposed a phenomenological model to scale the Rayleigh-B\'{e}nard convection. It hypothesizes existence of a turbulent boundary layer
near each plate, which is then shown to be convective, with a buoyancy dominated layer and two shear dominated layers. The leading-order
momentum balance and the convective logarithmic friction law of Tong and Ding \cite{TD19b} form the model equation set. Lie dilation
group of the equations were used to obtain the scaling relations. The $Nu$-$Ra$ and $Re$-$Ra$ scaling relations each has a single functional form
and is not a power law. Therefore the model predicts a single flow regime at high $Ra$ with no transition above $Ra>10^5$.
The $Nu$ scaling is close to $Ra^{1/3}$ for $Ra \sim 10^9$ to $10^{17}$, consistent with experimental and numerical results.
The model also predicts well the deviation from the $Ra^{1/3}$ scaling for $Ra<10^9$. 
The predicted single flow regime and the agreement between our prediction and the data for $Ra > 10^9$ and $Ra<10^9$
suggests  that the predicted $Nu$ scaling might hold at higher $Ra$ and will depart from $Ra^{1/3}$ for $Ra>10^{17}$.



\begin{acknowledgments}
This work was supported by the National Science Foundation under grant AGS-2054983.
\end{acknowledgments}

\vspace{-12pt}

\bibliography{../zpgbl/clemson1}
\end{document}